\documentclass[aps,prl,superscriptaddress,twocolumn]{revtex4-2}
\usepackage[latin9]{inputenc}
\usepackage{graphicx,epsf}
\usepackage{amssymb}
\usepackage{amsmath}
\usepackage{color,soul}
\usepackage{bm,bbm}
\usepackage{placeins}
\usepackage{verbatim}
\usepackage{textcomp}
\usepackage{esint}
\usepackage{array}
\usepackage{braket}
\usepackage{dsfont}
\usepackage{enumitem}

\usepackage{svg}
\usepackage{adjustbox}

\usepackage{tikz}

\definecolor{darkgreen}{rgb}{0,0.5,0}
\definecolor{darkblue}{rgb}{0,0,0.5}
\definecolor{purple}{rgb}{0.35,0,0.35}
\definecolor{orange}{rgb}{0.9,0.4,0}

\graphicspath{{./}{./figs/}{./pictures_main/}}

\begin{document}


\title{
Classical spin liquids from frustrated Ising models in hyperbolic space
}

\author{Fabian K\"ohler}
\affiliation{Institut f\"ur Theoretische Physik and W\"urzburg-Dresden Cluster of Excellence ct.qmat, Technische Universit\"at Dresden,
01062 Dresden, Germany}
\author{Johanna Erdmenger}
\affiliation{Institute for Theoretical Physics and Astrophysics and W\"urzburg-Dresden Cluster of Excellence ct.qmat, Julius-Maximilians-Universit\"at W\"urzburg, Am Hubland, 97074 W\"urzburg, Germany}
\author{Roderich Moessner}
\affiliation{Max-Planck-Institut f\"ur Physik komplexer Systeme and W\"urzburg-Dresden Cluster of Excellence ct.qmat, N\"othnitzer Str. 40, 01187 Dresden}
\author{Matthias Vojta}
\affiliation{Institut f\"ur Theoretische Physik and W\"urzburg-Dresden Cluster of Excellence ct.qmat, Technische Universit\"at Dresden,
01062 Dresden, Germany}

\date{\today}

\begin{abstract}
Antiferromagnetic Ising models on frustrated lattices can realize classical spin liquids, with highly degenerate ground states and, possibly, fractionalized excitations and emergent gauge fields. Motivated by the recent interest in many-body system in negatively curved space, we study hyperbolic frustrated Ising models. Specifically, we consider nearest-neighbor Ising models on tesselations with odd-length loops in two-dimensional hyperbolic space. For finite systems with open boundaries we determine the ground-state degeneracy exactly, and we perform extensive finite-temperature Monte-Carlo simulations to obtain thermodynamic data as well as correlation functions. We show that the shape of the boundary, constituting an extensive part of the system, can be used to control low-energy states: Depending on the boundary, we find ordered or disordered ground states. Our results demonstrate how geometric frustration acts in curved space to produce classical spin liquids.
\end{abstract}

\maketitle


The behavior of many-body systems is crucially influenced by symmetry and geometry of the underlying real-space lattice, with band degeneracies, band topology, as well as frustration and related interference effects being key examples. Interest in recent years has extended to lattices embedded in curved space, hyperbolic space in particular, for a number of reasons:
First, many condensed-matter paradigms established for Euclidean flat space are substantially modified in hyperbolic space, for instance band theory \cite{maciejko21,chen22,boettcher22} and critical behavior \cite{callan90,mnasri15,breuckmann20}.
Second, the holographic principle as realized by the AdS/CFT correspondence postulates a connection between interacting quantum field theories and gravity theories in higher dimensions \cite{witten98,maldacena99,gubser98}. As a generalization, it has also been postulated that gravity theories discretized on hyperbolic tilings of two-dimensional hyperbolic space are dual to spin chains at the boundary  \cite{axenidis14,brower21,basteiro22,dey24,erdmenger25}, a program known as `discrete holography'.
%
Third, hyperbolic systems have been discussed in quantum information theory as candidates to efficiently encode qubits using toric code \cite{breuckmann17}.

This paper is concerned with lattice models of interacting local moments in hyperbolic space. Defining such models has multiple aspects which differ from flat space: (i) The lattice geometry is dictated by the possibilities to discretize curved space in a periodic manner: This leads to distinct hyperbolic tilings \cite{mosseri82} leading to distinct lattice geometries. (ii) Boundaries and boundary conditions can influence bulk properties even in the large-system limit. (iii) The form of magnetic interactions can be influenced by the curvature of the underlying space; this aspect appears particularly relevant for vectorial (as opposed to Ising-type) local moments \cite{baek09}.

Previous work on hyperbolic magnetism has studied the two-dimensional ferromagnetic Ising model in some detail \cite{rietman92,krcmar08a,gendiar12,breuckmann20,asaduzzaman22}, which shows a mean-field-type transition into a low-temperature ferromagnetic phase. Other papers have studied the Ising model with competing second-neighbor interaction \cite{krcmar08b} as well as with random-bond disorder \cite{placke23}; and curvature as source of uniform frustration in continuum models has been discussed \cite{nelson83,mnasri15}.

In this paper, we focus on a different variety of hyperbolic magnetism, namely non-disordered frustrated lattice models. Given that frustration tends to suppress magnetic order, frustrated magnets herald many non-trivial forms of order and disorder. Prominent examples are various kinds of spin liquids, both classical and quantum, which often are characterized by fractionalized excitations and emergent gauge theories \cite{balents10,savary_rop17,fieldguide}. Notably, recent work has proposed realizations of the exactly solvable Kitaev model -- where spin liquidity is driven by exchange frustration -- on hyperbolic lattices \cite{dusel24,lenggenhager24}. However, spin liquids driven by \textit{geometric} frustration have not been considered to our knowledge, with the exception of a dimer-based wavefunction study in Ref.~\onlinecite{elser93}.
We therefore choose to investigate the antiferromagnetic nearest-neighbor (NN) Ising model
\begin{equation}
\mathcal{H} = J \sum_{\langle ij\rangle} \sigma_i \sigma_j
\label{hising}
\end{equation}
where $\sigma_i=\pm 1$ corresponds to spin-$1/2$ degrees of freedom.

The antiferromagnetic Ising model is known to generate classical spin liquids on various flat-space lattices, such as the triangular lattice  \cite{wannier50} and the pyrochlore lattice, the latter example being known as spin ice \cite{bramwell01,monopole08} which realizes the same statistical mechanics as water ice \cite{pauling35}. These classical spin liquids are characterized by the absence of magnetic order and an exponentially large number of lowest-energy states, implying a finite residual entropy density.

Here we study Ising spins placed on two-dimensional hyperbolic lattices, assuming an interaction of the form \eqref{hising} irrespective of the curvature of the underlying space. We demonstrate that classical spin liquids can be realized in hyperbolic space, and we elucidate on the similarities to and differences from their flat-space counterparts. A remarkable result is that the boundary shape can be used to \textit{control} the low-temperature magnetic phases.


\paragraph{Hyperbolic tessellations and geometric frustration.---}
Discretizations of curved space lead to the concept of tesselations: A uniform tessellation $\{p,q\}$ denotes a two-dimensional space-filling lattice of regular $p$-sided polygons where $q$ polygons meet at each lattice vertex. Such tessellations can naturally be embedded on two-dimensional surfaces of constant curvature $\rho$ \cite{boettcher22,boyle20}. The sign of the \textit{characteristic} $\chi = 1/2 - 1/p -1/q$ distinguishes between spherical ($\chi < 0$), flat ($\chi = 0$) and hyperbolic ($\chi > 0$) tessellations, embedded on the eponymous space.
Geometric frustration requires $p$ to be odd. To generate maximal frustration, we focus on hyperbolic tessellations that are composed of triangles, $p=3$. Since the simplest realization of $\chi>0$ is given by $q=7$, we focus on the $\{3,7\}$ tessellation for most of this work.

\begin{figure}[!tb]
\centering
\includegraphics[width=0.95\columnwidth]{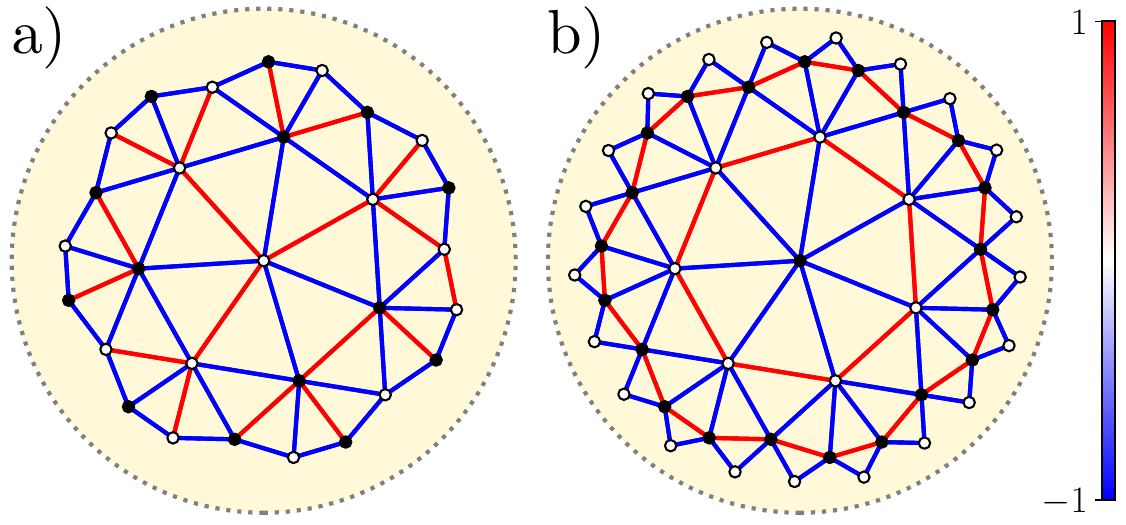}
\caption{
Snapshot low-temperature configurations of antiferromagnetic Ising models on $\{3,7\}$ lattices. Red/blue bonds indicate parallel/antiparallel spins on the respective bond.
(a) System with type-A boundary with $N=29$ spins, showing a spin-liquid configuration with antiferromagnetic alignment along the boundary.
(b) Type-B boundary with $N=50$ spins, showing an ordered ferrimagnet.
}
\label{fig:config}
\end{figure}

Uniform hyperbolic tessellations represent infinite lattices; they can be understood as limits of finite hyperbolic uniform tilings. These finite lattices have unique features that distinguish them from their planar counterparts: The number of sites grows exponentially with linear system size, leading to strong boundary contributions.
Implementing periodic boundary conditions is not straightforward: Their possible unit cells are restricted by the Gauss-Bonnet theorem and involve the solution of lattice-specific diophantine equations \cite{boettcher22,breuckmann20}; we therefore choose to work with open boundary conditions. There are a multitude of ways to construct finite lattices, themselves leading to different shapes of the lattice boundary. We choose the iterative algorithm of Refs.~\onlinecite{basteiro22,boyle20}, described in more detail in the supplemental material (SM) \cite{suppl}. A sample lattice is shown in Fig.~\ref{fig:config}(a) as a planar graph using Poincar\'e projection; we refer to this type of boundary as type-A. Other boundary shapes will be discussed below.


\paragraph{Thermodynamics.---}
We study the thermodynamic behavior of the hyperbolic Ising model using Monte-Carlo (MC) simulations using Metropolis single-spin updates combined with parallel tempering, for details see SM \cite{suppl}. Simulations for large systems are plagued by low-temperature freezing, due to the lack of multi-spin updates in our algorithm \cite{clusternote}, such that we have restricted our simulations to systems with $N\lesssim 2000$ spins. 

\begin{figure}[tb!]
\includegraphics[width=\columnwidth]{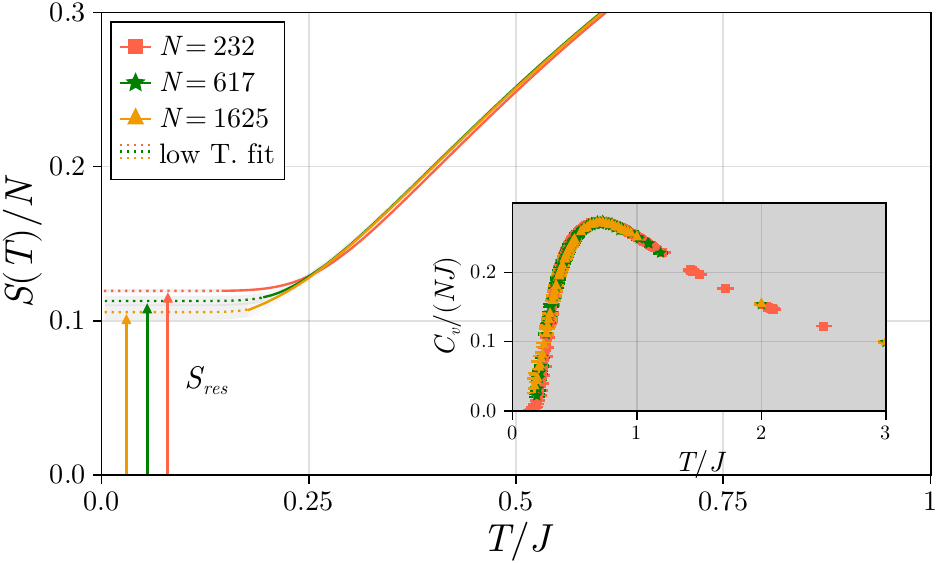}
\caption{
Entropy (main) and specific heat (inset) per site as function of temperature for $\{3,7\}$ type-A systems. The entropy, obtained from integrating the specific heat, is used to extract a residual entropy, for details see text.
}
\label{fig:S_vs_T}
\end{figure}

Results for entropy and specific heat of type-A systems are shown in Fig.~\ref{fig:S_vs_T}; a snapshot low-$T$ configuration is in Fig.~\ref{fig:config}(a). The specific heat $C_V(T)/N$ displays a broad maximum around $T/J \approx 0.6$ whose height and shape is essentially independent of system size. At low temperatures we get enlarged statistical errors because of spin freezing. The energy $E(T)/N$, shown in the SM \cite{suppl}, reaches a constant upon cooling below $T/J\lesssim 0.1$; this constant approaches $(-J)$ with increasing system size, in agreement with the analytical result obtained below.

The finite-temperature corrections to the $T=0$ values of energy and specific heat appear to be of thermally activated form. This is plausible given that the smallest finite energy cost of a spin flip from a ground-state configuration in a frustrated Ising model is $2J$; this applies in particular to a finite fraction of the boundary spins, Fig.~\ref{fig:config}(a). We can fit the thermodynamic quantities accordingly, e.g.
\begin{equation}
    \frac{E(T) - E(0)}{N} = \frac{\Delta}{\gamma \exp(\Delta/T) +1}
\end{equation}
with the gap $\Delta=2J$ and $\gamma$ the (unknown) ratio of the number of ground and excited states.

We use the specific heat, with a low-temperature fit as described, to compute the entropy according to
\begin{equation}
    S(T) = N\ln 2 - \int_T^\infty \! dT'\,\frac{C_V(T')}{T'}
\end{equation}
in units where $k_B=1$, using that $S(T\to\infty)=N\ln 2$. In practice, we integrate up to $T'/J=100$ \cite{expansionnote}.
The results for $S(T)$ in Fig.~\ref{fig:S_vs_T} for type-A systems indicate a finite residual entropy density, $s_{\rm res}=S(T\to 0)/N$, of order $0.1$ whose dependence on system size is discussed below.

\begin{figure}[tb!]
\includegraphics[width=\columnwidth]{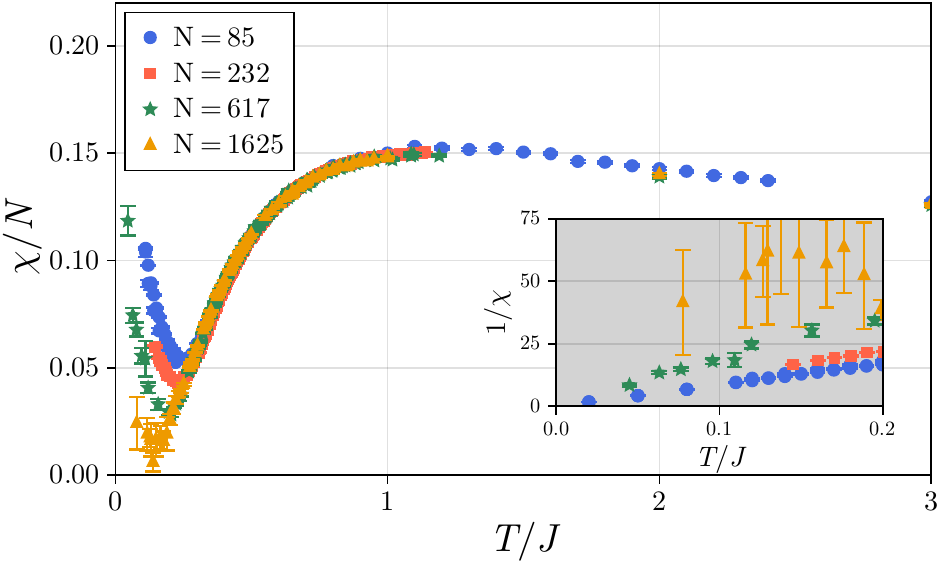}
\caption{
Uniform magnetic susceptibility $\chi$ as function of temperature $T$ for type-A systems;
the inset shows $1/\chi$ at low $T$.
}
\label{fig:zerofieldChi}
\end{figure}

We also use MC to compute the zero-field uniform magnetic susceptibility $\chi(T)$, with results shown in Fig.~\ref{fig:zerofieldChi}. We observe a very broad maximum in $\chi(T)/N$ around $T/J=1.2$. Below $T/J \sim 0.3$ data for different system sizes strongly differ, each obeying a low-temperature Curie law, $\chi \sim 1/T$, with the prefactor decreasing with system size. As detailed below and in SM \cite{suppl}, the reason is that the ground-state manifold contains states with non-zero magnetization, but the magnetization distribution in this manifold has a non-trivial scaling with system size.

\begin{figure}[!tb]
\includegraphics[width=\columnwidth]{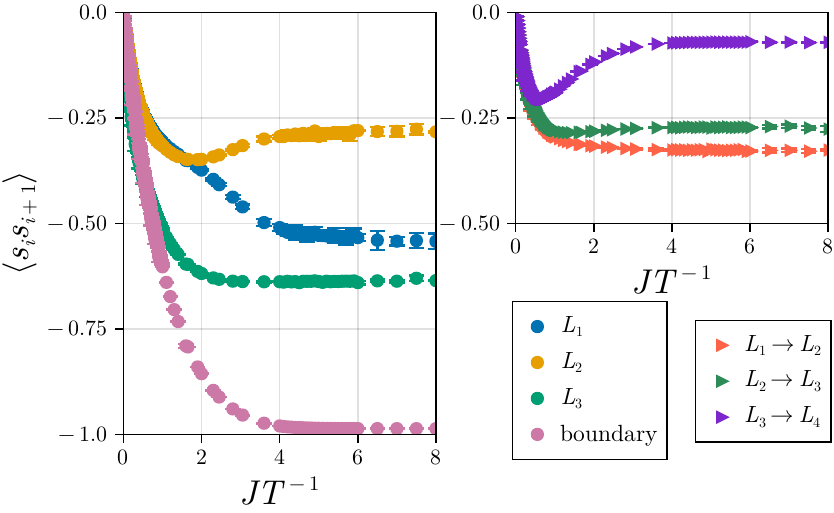}
\caption{
(a) Average NN spin correlations in each layer $L_i$ as function of $1/T$ of the $N=617$ type-A lattice (four layers); the fourth layer being the boundary.
(b) Average NN spin correlations between neighboring layers.
}
\label{fig:avg_cnn}
\end{figure}

\paragraph{Correlation functions.---}
As the lattices are built up in layers, it is useful to analyze the spin configurations layer by layer, where each layer can be interpreted as a one-dimensional (1D) chain. Numerical data for low-temperature NN spin correlations are shown in Fig.~\ref{fig:avg_cnn} for a moderately sized lattice, where the data have been averaged over all NN bonds of one layer (panel a) or over all NN bonds connecting two neighboring layers (panel b). In addition, thermal averages for each individual bond of a small lattice are shown in Fig.~\ref{fig:config}(b).
For these type-A systems, the boundary layer is least frustrated due to the lack of outer links; at low temperatures it essentially orders into an antiferromagnetic chain pattern: depending on the system size, the boundary layer has an even (odd) number of sites, leading to zero (one) frustrated intralayer bond in the ground state, respectively; this can also be nicely seen in the snapshot shown in Fig.~\ref{fig:config}(a). Fig.~\ref{fig:avg_cnn} shows that frustration is larger for the inner layers, and correlations within the layers tend to be stronger than between layers, except for the inner core of the system. The temperature evolution of the correlations and their dependence on distance are analyzed in SM \cite{suppl}.

\begin{figure}[!tb]
\includegraphics[width=\columnwidth]{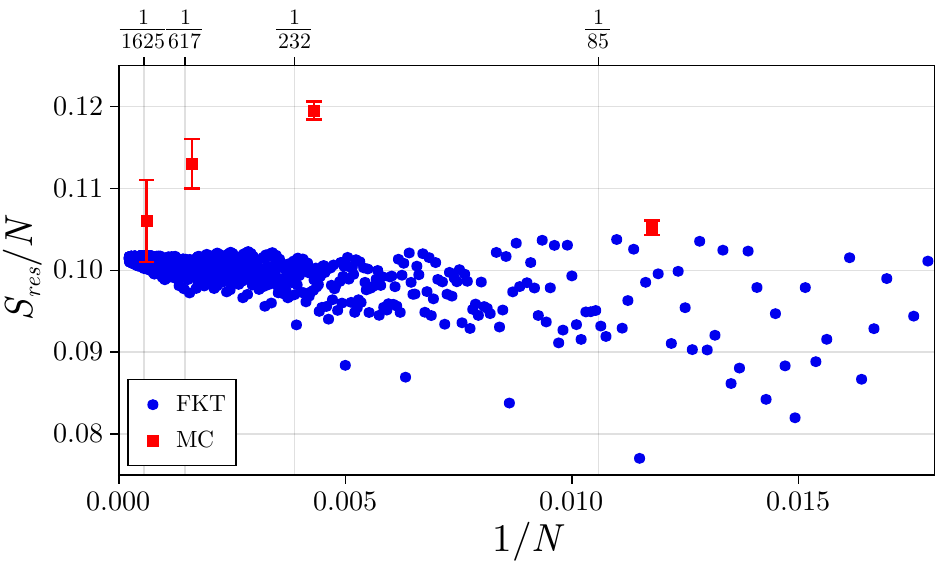}
\caption{
Residual entropy per spin of $\{3,7\}$ type-A systems as function of inverse system size, obtained from the FKT algorithm; MC results are given for comparison.
}
\label{fig:resEntr_FKT+MC}
\end{figure}

\paragraph{Ground-state manifold.---}
The manifold of lowest-energy states of the Ising model under consideration can be obtained semi-analytically, using the Fisher-Kasteleyn-Temperley (FKT) algorithm \cite{temperley61,kasteleyn61}. For Ising models on planar graphs with odd plaquette loop lengths, there is at least one frustrated bond per plaquette. A ground-state configuration can be mapped onto a (perfect) dimer covering on the dual graph, provided that the graph can host a dimer covering at all and its number of plaquettes is even. The FKT algorithm enables to efficiently count such coverings via the skew-symmetric adjacency matrix $A'$ of the directed dual graph, for details see SM \cite{suppl}. The residual entropy is then given by
\begin{equation}
    S_{\rm res} = \ln(2)+\frac{1}{2} \ln (\det A'),
\label{s_fkt}
\end{equation}
which we have used to compute $S_{\rm res}$ for system sizes up to $N=6000$. 
The corresponding numerical results for $\{3,7\}$ type-A systems are shown in Fig.~\ref{fig:resEntr_FKT+MC}; here we have also included data for systems with partially completed layers. MC data for $s_{\rm res}$ are shown for comparison; these are consistent with the FKT data within error bars for cases where the same system size is covered. We note that the $N=232, 617$ systems feature an odd number of plaquettes such that the FKT algorithm is not directly applicable; in that case the frustrated bond on the boundary provides an additional entropy of order $\mathcal{O}(\ln(N_{\rm bnd}))$, with $N_{\rm bnd}$ the number of boundary sites. Despite significant finite-size effects, the residual entropy clearly converges to a finite value for $N\to\infty$, which we estimate as $s_{\rm res}\equiv S_{\rm res}/N=0.102(2)$.

Compared to the planar triangular-lattice Ising antiferromagnet, where single spin flips within the ground-state manifold are possible, the minimum flippable motif in the hyperbolic $\{3,7\}$ case is pair flips. This is related to the coordination number being odd.

The dimer mapping can also be used to derive the following expression for the ground-state energy \cite{suppl}
\begin{equation}
 \frac{E_{GS}}{J \, N} = \frac{2 \lceil N_P / 2 \rceil - N_E}{N},
\end{equation}
where the number of vertices ($N$), edges ($N_E$) and plaquettes ($N_P$) of the lattice are related by Euler's characteristic for planar graphs, $N-N_E+N_P = 1$; this differs by unity from the literature since our definition of plaquettes excludes the outside. For our hyperbolic lattice we can obtain closed expressions for $N$, $N_E$, and $N_P$ based on its iterative construction. This eventually leads to \cite{suppl}
\begin{equation}
    \frac{E_{GS}}{J \, N} \approx -1 + \frac{2\sqrt{5}(1+\delta_n)}{7 (1+\sqrt{5})} \left( \frac{3+\sqrt{5}}{2} \right )^{-n}
\end{equation}
valid for large $n$ where $n$ denotes the number of lattice layers, and $\delta_n$ is one if $n$ is divisible by three and zero otherwise. The ground-state energy hence approaches $(-JN)$ for $N\to\infty$, in agreement with the numerical results in the SM \cite{suppl}.


\paragraph{Classical spin liquid.---}
Our analytical and numerical data for the antiferromagnetic Ising model on type-A hyperbolic lattices are consistent with the presence of a classical spin liquid: This is indicated by the absence of a thermal phase transition, by the absence of global long-range order at $T=0$, and by a finite residual entropy density.
The realization of a spin liquid in the present setting is non-trivial, as the open boundary is less frustrated than the bulk. Given that the fraction of boundary sites is finite in the $N\to\infty$ limit, $N_{\rm bnd}/N \to 0.62$ for $\{3,7\}$ type-A hyperbolic tilings \cite{suppl}, the system under consideration is only partially frustrated. This is reflected in the residual entropy, $s_{\rm res}=0.102(1)$, being smaller than that for the planar triangular-lattice Ising model, $s_{\rm res}=0.32306$ \cite{wannier50}.

For comparison, we have repeated the MC simulations for the ferromagnetic Ising model on the same lattices. Here we observe clear signatures of a continuous thermal phase transition into a ferromagnetic state, consistent with earlier results \cite{rietman92,krcmar08a,gendiar12,breuckmann20,asaduzzaman22},
for details see SM \cite{suppl}.

\begin{figure}[!tb]
\includegraphics[width=\columnwidth]{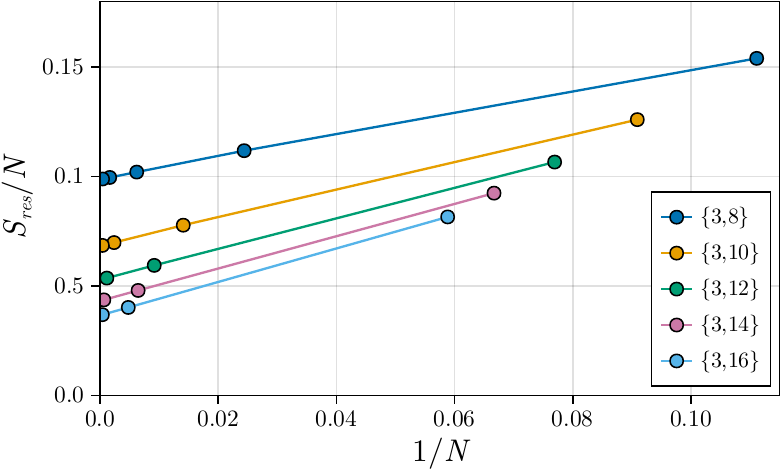}
\caption{
Residual entropy per spin of $\{3,q\}$ type-A systems with $q\geq8$ as function of inverse system size, obtained from the FKT algorithm; lines are guide to the eye.
}
\label{fig:resentr_3q}
\end{figure}

To verify the universality of the spin-liquid behavior in the antiferromagnetic case, we have extended our study to frustrated tilings $\{3,q\}$ with $q>7$, again with type-A boundaries. Results for the residual entropy obtained from the FKT algorithm are displayed in Fig.~\ref{fig:resentr_3q}, unambiguously showing a finite $s_{\rm res}$ in the large-system limit. The value of $s_{\rm res}$ decreases with increasing $q$, consistent with the fact that the fraction of (unfrustrated) boundary sites increases with $q$.

\begin{figure}[!tb]
\centering
\includegraphics[width=0.95\columnwidth]{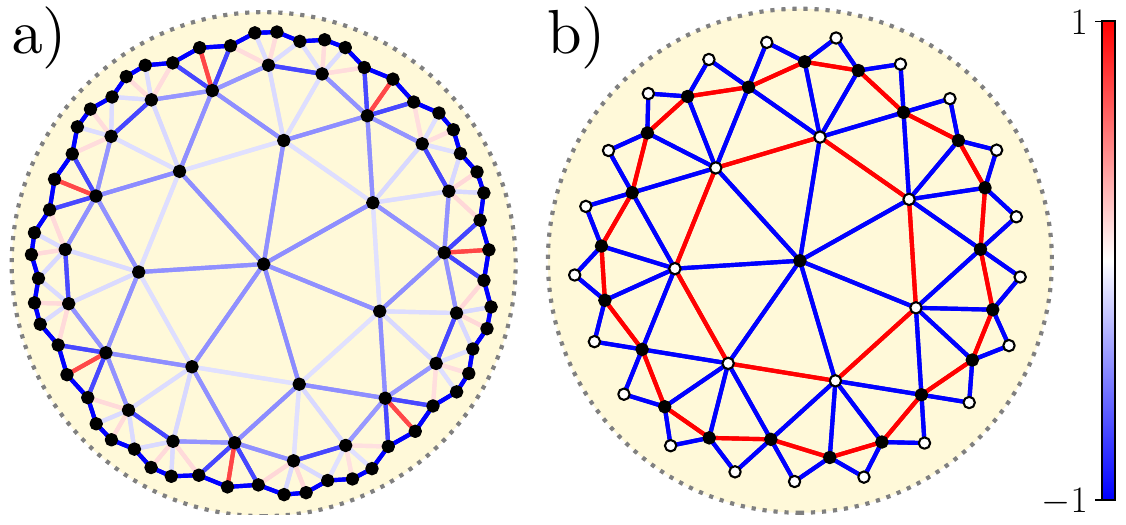}
\caption{
NN bond correlations $\langle\sigma_i\sigma_j\rangle$ (color-coded) for $\{3,7\}$ systems at $T\!=\!0.2$.
(a) Type-A with $N=85$ in the spin-liquid regime.
(b) Type-B with $N=50$, indicating ferrimagnetic order.
}
\label{fig:TD_avg}
\end{figure}

\paragraph{Boundary control.---}
We have also studied hyperbolic $\{3,7\}$ tilings with different boundary geometry. A type-B lattice is obtained by adding one site, i.e. one outer triangle, to each boundary bond of a type-A lattice. This construction generates a family of type-B tilings, Fig.~\ref{fig:config}(b). Our numerical data for type-B systems indicate that they settle into an ordered ferrimagnetic ground state, with ferromagnetic order in each layer and antiferromagnetic alignment from layer to layer, Fig.~\ref{fig:TD_avg}. The reason is energetic: Type-B systems have more boundary bonds than type-A systems, with a zigzag boundary shape, such that perfect antiferromagnetic order along the boundary coincides with ferromagnetic intralayer order, thus tipping the balance towards ordered states. More details on the behavior of type-B systems are in the SM \cite{suppl}.

\paragraph{Summary and outlook.---}
We have shown that the antiferromagnetic Ising model on $\{3,q\}$ tilings realizes hyperbolic classical spin liquids, which we have characterized using thermodynamic and ground-state analyses. Our results indicate the presence of a well-defined large-system limit with open boundary conditions where the residual entropy density approaches a unique non-zero value.
We have also demonstrated that spin liquidity can be controlled via the boundary: By changing the boundary shape from type-A to type-B, we can tune the energetic balance towards ordered ferrimagnetic states; the latter keep the spin-liquid manifold as low-lying excited states. Studying further boundary shapes \cite{hypertiling} is subject of ongoing work.

Among the directions for future research are the investigation of quantum effects towards triangular hyperbolic quantum spin liquids -- this suggests for instance to consider quantum dimer models on the present lattices -- as well as the study of classical spin liquids in hyperbolic vector spin models. More broadly, open questions concern how lattice frustration manifests in different curved-space geometries, e.g., with black holes included \cite{dey24,chen23}, and whether the spin liquids found here have implications for gravitational theories.
We note that the finite residual entropy found here appears as an essential ingredient of the AdS/CFT correspondence: It was shown that the residual entropy of the Sachdev-Ye model \cite{SachdevYe} matches the entropy associated to the AdS$_2$ throat arising in Reissner-Nordstr\"om charged black hole geometries \cite{sachdev15} -- a result which triggered intense research establishing holographic dualities for the Sachdev-Ye-Kitaev (SYK) model \cite{SYK}. Since the Sachdev-Ye model is a spin model, our results raise the possibility that a similar duality may be realized between spin chains and antiferromagnetic Ising models. A first check here will be to compare  correlation functions for spin chains at the boundary of hyperbolic spaces, as calculated e.g. in Ref.~\onlinecite{DiGiulio}, with the correlation functions found here. Moreover, our results emphasize the strong influence of the boundary conditions on the ground-state degeneracy and thus on the possible presence of a gravity dual.


\acknowledgments

We thank P. Basteiro, P. M. C\^onsoli, J. Karl, R. Meyer, B. Placke, and Y. Thurn
for helpful discussions.
Financial support from the Deutsche Forschungsgemeinschaft through SFB 1143 (project-id 247310070), SFB 1170 (project-id 258499086), and the W\"urzburg-Dresden Cluster of Excellence on Complexity and Topology in Quantum Matter -- \textit{ct.qmat} (EXC 2147, project-id 390858490) is gratefully acknowledged.



\begin{thebibliography}{}
\bibitem{clusternote}
Standard cluster update schemes do not work well in MC simulations on hyperbolic lattices because of the large boundary fraction of sites.
\bibitem{expansionnote}
Alternatively, a high-temperature expansion can be used to obtain the exact behavior of $C_V(T)$ in the high-$T$ limit.
\end{thebibliography}
\end{document}



\title{
Supplemental material for:\\
Classical spin liquids from frustrated Ising models in hyperbolic space
}

\author{Fabian K\"ohler}
\affiliation{Institut f\"ur Theoretische Physik and W\"urzburg-Dresden Cluster of Excellence ct.qmat, Technische Universit\"at Dresden,
01062 Dresden, Germany}
\author{Johanna Erdmenger}
\affiliation{Institute for Theoretical Physics and Astrophysics and W\"urzburg-Dresden Cluster of Excellence ct.qmat, Julius-Maximilians-Universit\"at W\"urzburg, Am Hubland, 97074 W\"urzburg, Germany}
\author{Roderich Moessner}
\affiliation{Max-Planck-Institut f\"ur Physik komplexer Systeme and W\"urzburg-Dresden Cluster of Excellence ct.qmat, N\"othnitzer Str. 40, 01187 Dresden}
\author{Matthias Vojta}
\affiliation{Institut f\"ur Theoretische Physik and W\"urzburg-Dresden Cluster of Excellence ct.qmat, Technische Universit\"at Dresden,
01062 Dresden, Germany}

\date{\today}

\maketitle


\section{Hyperbolic tessellations}
\label{sec:hyper}

Tesselations provide the general concept to discretize (hyperbolic) space: In two dimensions, a uniform tessellation $\{p,q\}$ is a space-filling lattice of $p$-sided polygons, with $q$ polygons meeting at each vertex. The sign of the characteristic $\chi = 1/2 - 1/p -1/q$ determines whether the tesselation is embedded into a spherical ($\chi < 0$), flat ($\chi = 0$), or hyperbolic ($\chi > 0$) space with constant curvature $\rho$ \cite{boettcher22}. Our study is concerned with hyperbolic tessellations that are composed of triangles, $p=3$ and $q\geq 7$. 
All hyperbolic tessellations are infinite lattices and can be represented as planar graphs, commonly visualized using the Poincar\'e disc projection. In such a projection the lattice vertices (and edges) are mapped to the interior of a unit disk $\{(x,y), r^2 = x^2 +y^2 < 1\}$ with hyperbolic metric \cite{boettcher22}
\begin{equation}
 ds^2 = \frac{4}{\rho^2} \frac{dx^2 + dy^2}{(1-r^2)^2}.
\end{equation}
%
We note that, in curved space, the relation between $\chi$, $\rho$, and the bond length $a$ is fixed, such that one cannot scale the lattice as in planar space \cite{boettcher22}.

\subsection{Translations and momentum space}
The underlying symmetry group of all uniform tessellation is the \textit{triangle group} $\Delta(p,q,2)$ which consist of vertex-centered and plaquette-centered rotations as well as inversions. For flat tesselations we know that there exists a subgroup of commuting translations so that we can define a unit cell or super cell for the lattice with periodic boundary conditions (PBC). Using Bloch's theorem we can introduce momentum space which also forms a periodic tessellation of planar space. Both the real-space and momentum-space cells with PBC form a \textit{1-torus} independent of cell size.

In contrast, in hyperbolic tessellations the translational subgroups are in general non-commuting, and the choice of periodic super cells is limited by the Gauss-Bonnet theorem \cite{boettcher22,breuckmann20}. Not only do they form tori of genus higher than one, the genus of the cells can become arbitrarily large. The number of independent momentum-space vectors is twice the genus of the periodic super cell and by that not fixed for hyperbolic tessellations.
Furthermore, finding a super cell with periodic boundary condition involves the solutions of diophantine equations, which makes a systematic approach to size scaling impractical.
\subsection{Finite lattices and open boundaries}
For this work we therefore focus on finite lattices with open boundary conditions (OBC). Doing so allows us to be more flexible with system size compared to PBC and to investigate the effects of the boundary on thermodynamic properties. There are a multitude of ways to construct a finite lattice, i.e., a hyperbolic tiling, which lead to different shapes of the lattice boundary. Often, the construction iteratively grows the lattice in layers of increasing radius.
One important feature, which is common to all constructions and which distinguishes OBC systems in hyperbolic space from their flat-space counterparts, is that the total number of sites, $N$, and the number of sites on the boundary, $N_{\rm bnd}$, grow exponentially with linear system size, i.e., with the number of layers. The number of boundary sites approaches some fixed (and sizeable) fraction of $N$ in the limit $N\to\infty$. The leads to strong boundary contributions to all physical observables even in the large-system limit, and it restricts the linear system sizes amendable to numerical simulations.

We specifically use the following algorithm to iteratively grow a lattice \cite{basteiro22}:
\begin{enumerate}
 \item set a vertex of a $\{p,q\}$ tesselation as the origin of the flake (layer $L_0$);
 \item identify all vertices at the current boundary, i.e., with coordination less than $q$;
 \item then insert the missing neighbors of that vertex and close the newly started plaquettes with loop length $p$ (layer $+1$);
 \item repeat steps 2 and 3 until layer $L_n$ is reached.
\end{enumerate}
This construction yields a ``smooth'' boundary, compatible with the FKT algorithm described below. The concrete inflation rules for step 3 will be described now.

\subsection{Inflation rules and recursion for the type-A $\{3,7\}$ tiling}
\label{sec:infla}

We build up the grid from a central vertex, $L_0$, at the center of the Poincar\'e disc. To construct the next layer, $L_{i+1}$, we identify all bonds missing from layer $L_i$ and add new grid vertices accordingly. Then we insert bonds such that triangular plaquettes are formed, Fig.~\ref{fig:inflation}. Starting from layer one, $L_1$, we track number of missing edges for each vertex of the boundary (exposure of the vertex) as a string of numbers.
\begin{figure}[!tb]
\begin{tikzpicture}[
L1node/.style={circle, draw=blue!60, fill=blue!5, very thick, minimum size=7mm},
L2node/.style={circle, draw=red!60, fill=red!5, very thick, minimum size=5mm},
Plaq/.style ={black, very thick, fill=gray!10}
]

\draw[Plaq] (-1,2) -- (0,3) -- (0,0)-- (-1,2);
\draw[Plaq] (0.8,2) -- (0,3) -- (0,0)-- (0.8,2);

\draw[Plaq] (0.8,2) -- (0,0) -- (2,0)-- (0.8,2);

\draw[Plaq] (2,0) -- (0.8,2) -- (1.5,3)-- (2,0);
\draw[Plaq] (2,0) -- (2.5,3) -- (1.5,3)-- (2,0);
\draw[Plaq] (2,0) -- (2.5,3) -- (3,2)-- (2,0);

\draw[Plaq] (2,0) -- (4,0) -- (3,2)-- (2,0);

\draw[Plaq] (4,0) -- (3,2) -- (3.5,3)-- (4,0);
\draw[Plaq] (4,0) -- (4.5,3) -- (3.5,3)-- (4,0);
\draw[Plaq] (4,0) -- (4.5,3) -- (5,2)-- (4,0);

\draw[dashed, very thick] (-1,2) -- (-1.7,2);
\draw[dashed, very thick] (-1,2) -- (-1.7,1.3);

\draw[dashed, very thick] (5,2) -- (5.7,2);
\draw[dashed, very thick] (5,2) -- (5.7,1.3);

\node[L2node] at (-1,2) {3};
\node[L2node] at (0,3) {4};
\node[L2node] at (0.8,2) {3};
\node[L2node] at (1.5,3) {4};
\node[L2node] at (2.5,3) {4};
\node[L2node] at (3,2) {3};
\node[L2node] at (3.5,3) {4};
\node[L2node] at (4.5,3) {4};
\node[L2node] at (5,2) {3};

\draw[dashed, very thick] (-0.8,0) -- (0,0);
\draw[dashed, very thick] (0,0) -- (0.7,-0.7);
\draw[dashed, very thick] (0,0) -- (-0.7,-0.7);

\draw[dashed, very thick] (2,0) -- (2,-0.7);

\draw[dashed, very thick] (4,0) -- (4.8,0);
\draw[dashed, very thick] (4,0) -- (3.3,-0.7);

\draw[very thick] (0,0) -- (3,0) -- (4,0);
\node[L1node] at (0,0) {3};
\node[L1node] at (2,0) {4};
\node[L1node] at (4,0) {4};

\node[blue!60, anchor=west] at (5.5,0) {$L_i$};
\node[red!60,anchor=west] at (5.5,2.4) {$L_{i+1}$};

\draw[blue!60, thick, dashed] (-1.5,0.8) -- (5.5,0.8) -- (5.5,-0.8) -- (-1.5,-0.8) -- (-1.5,0.8);

\draw[red!60, thick, dashed] (-1.5,3.6) -- (5.5,3.6) -- (5.5,1.2) -- (-1.5,1.2) -- (-1.5,3.6);

\end{tikzpicture}
\caption{Inflation rules for one iteration for the $\{3,7\}$ tiling. Nodes are labelled by their respective exposure.}
\label{fig:inflation}
\end{figure}
By applying the algorithm to singular vertices and minding the connections to neighboring vertices, we can see that solely depending on the exposure each vertex ``spawns'' new vertices in the next layer. The inflation rules can then be expressed in terms of strings of integers as
\begin{equation}
    (3) \rightarrow (34), \; (4) \rightarrow (344)
\end{equation}
for layers one and onward we only encounter exposures of $3$ and $4$. For the total string of the first layers this gives,
\begin{equation}
    (4 \dots) \rightarrow (344 \dots) \rightarrow (34344344 \dots) \rightarrow
\end{equation}
where the dots indicate seven repetitions because of the sevenfold rotational symmetry of the lattice. We dub the number of vertices with exposure $3,4$ in layer $i$ as $X_{3,4}^{i}$ respectively. From the inflation rules we can infer the recursive linear relation
\begin{equation}
    \begin{pmatrix}
    X_3^{i+1} \\
    X_4^{i+1}
    \end{pmatrix}
    =
        \begin{pmatrix}
    1 & 1 \\
    1 & 2
    \end{pmatrix}
        \begin{pmatrix}
    X_3^{i} \\
    X_4^{i}
    \end{pmatrix}
\end{equation}
which is reminiscent of the definition of Fibonacci or Lucas numbers. The starting value for the first layer is $(0,7)$. We could also incorporate the initial vertex with exposure $7$ by defining $X_7^0 = 1$ and adjusting the recursion matrix.

The number of vertices, $N^{n}$, in each layer is the sum of its exposure counts, and the number of vertices of a lattice up to layer $n$ is simply the sum of all layers up to $n$ and the initial vertex:
\begin{equation}
    N^{n} = 1 + \sum_{i=1}^n (X_3^i + X_4^i).
\end{equation}
For the number of edges we can employ the handshaking lemma from graph theory. The number of edges will be half of the sum of all vertex degrees \cite{euler1736}. A bulk vertex has degree $7$, and a boundary vertex has its degree decreased by its exposure. The expression for the number of edges is given as
\begin{equation}
    N_E^{n} = \frac{1}{2} \left (7 N^n - 3 X_3^n - 4 X_4^n \right ),
    \label{eq:exp-E}
\end{equation}
and furthermore we can calculate, using Euler's characteristic, the number of plaquettes (only interior) as
\begin{equation}
    N_P^{n} = 1 - N^n + N_E^n.
    \label{eq:euler-P}
\end{equation}
This allows us to compute the relevant graph quantities from the inflation rules. An explicit expression for the exposure counts is obtained by diagonalizing the inflation matrix, resulting in
\begin{equation}
    X^n = \frac{7}{10} (5+\sqrt{5}) \lambda_+^{n-1} v_+ + \frac{7}{10} (5-\sqrt{5}) \lambda_-^{n-1} v_-,
\end{equation}
with eigenvectors $v_\pm = (-1/2 \pm \sqrt5/2,1)$ and the respective eigenvalues $\lambda_\pm = (3 \pm \sqrt{5})/2$ both greater zero. For large system size only the larger of the two eigenvalues enters.
The number of vertices can be expressed in closed form as
\begin{equation}
 N^n = \frac{7\sqrt{5}}{5} (\phi \lambda_+^n + \phi^{-1} \lambda_-^n) -6
 \label{eq:explicit-NV}
\end{equation}
with $\phi=(\sqrt{5}+1)/2$ the golden ratio, from which one can derive closed forms of $N_E^n, N_P^n$ using Eqs.~\eqref{eq:exp-E} and \eqref{eq:euler-P}.

\begin{figure}[tb]
\includegraphics[width=\columnwidth]{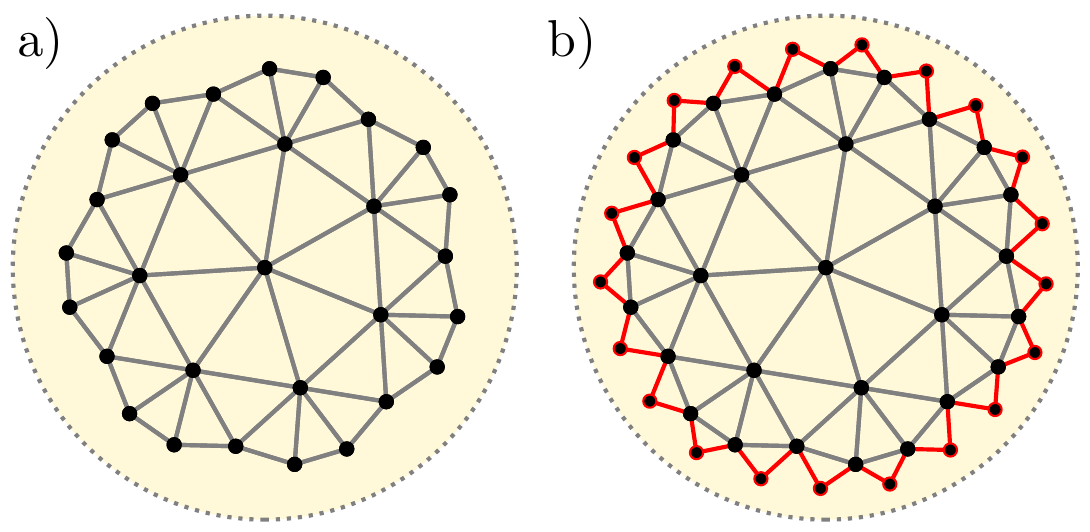}
\caption{
(a) $\{3,7\}$ lattice with $n=2$ layers and type-A boundary.
(b) The same lattice, now with type-B boundary, obtained from decorating each edge bond by an additional triangle (red).
}
\label{fig:spiked_boundary}
\end{figure}

We can use these results to extract the fraction of boundary sites, $N_{\rm bnd}/N = 1-N^{n-1}/N^n$. For the type-A $\{3,7\}$ lattices at hand this approaches $(\sqrt{5}-1)/2 = 0.618$ in the large-system limit. For $\{3,q\}$ lattices this ratio grows with $q$ and reaches, e.g., a value close to $0.9$ for $\{3,14\}$.

\subsection{Type-B boundaries}
\label{sec:typeblatt}

To study the influence of the boundary shape on physical properties, we consider -- in addition to the type-A lattices described above -- type-B lattices which are obtained by adding new plaquettes to the boundary, specifically one triangle per edge bond of the type-A lattice, see Fig.~\ref{fig:spiked_boundary}.

Starting from a type-A system with $n$ layers, this construction increases the numbers of vertices and plaquettes each by $N^n$ as well as the number of edges by $2N^n$. As a result, a type-B lattice always has an even number of plaquettes.


\section{Ground states of the frustrated Ising model}
\label{sec:gs}

The antiferromagnetic Ising model on a $\{3,7\}$ tessellation has one frustrated bond per plaquette with leads to an extensive ground-state degeneracy. To minimize the ground-state energy means to find the minimal number of frustrated bonds possible. This leads to a ground-state energy of
\begin{equation*}
    E_{GS}/J = 2 \times \# \text{frustrated bonds} - \# \text{total bonds}
\end{equation*}
For a planar graph of connected Ising variables the possible ground states can be mapped to a covering of dimers on the dual graph, a $\{7,3\}$ tessellation. For a finite lattice this comes with two caveats, (i) the number of plaquettes must be even and (ii) there are shapes that cannot host a dimer covering, Fig.~\ref{tikz:dimer_cover_Ex}. This problem can be lifted by considering the outside as an additional plaquette or allowing double occupancies for the dimers. In graph theory the dimer coverings are also referred to as perfect matchings. Their number can be calculated directly from the Hafnian of the adjacency matrix, $\textrm{haf}(A)$. It is not feasible to calculate the Hafnian for large matrices as its computation time scales like $\mathcal{O}(n^{3}2^{n})$. Using the Fisher-Kasteleyn-Temperley (FKT) algorithm \cite{temperley61,kasteleyn61} we can instead calculate the Pfaffian of a related skew-symmetric matrix in polynomial time.
\begin{figure}[tb!]
 \begin{tikzpicture}
\begin{scope}[]
\node[rounded corners=2pt, inner sep=4pt, fill=red!50] at (0,2) {a};

\draw (0.5,1) -- (0,0) -- (1, 0) -- (2,0) -- (1.5,1) -- (1,0) -- (0.5, 1) -- (1.5,1) -- (1,2) -- (0.5,1) ;
\draw (2,0) -- (2.5,1) -- (1.5,1) -- (1,2) -- (2,2) -- (1.5,1);

\draw[red,ultra thick] (0.5,0.33) -- (1,0.66);
\draw[red,ultra thick] (1.5,0.33) -- (2,0.66);
\draw[red,ultra thick] (1,1.33) -- (1.5,1.66);

\shade[ball color=blue] (0.5,0.33) circle (0.1);
\shade[ball color=blue] (1.5,0.33) circle (0.1);
\shade[ball color=blue] (1,0.66) circle (0.1);
\shade[ball color=blue] (2,0.66) circle (0.1);
\shade[ball color=blue] (1,1.33) circle (0.1);
\shade[ball color=blue] (1.5,1.66) circle (0.1);
\end{scope}

\begin{scope}[xshift=3cm]
\node[rounded corners=2pt, inner sep=4pt, fill=red!50] at (0,2) {b};

\draw (0.5,1) -- (0,0) -- (1, 0) -- (2,0) -- (1.5,1) -- (1,0) -- (0.5, 1) -- (1.5,1) -- (1,2) -- (0.5,1) ;
\draw (2,0) -- (1.5,1) -- (1,2) -- (2,2) -- (1.5,1);

\draw[red,ultra thick] (0.5,0.33) -- (1,0.66);
\draw[red,ultra thick, dashed] (1.5,0.33) to[out=20,in=-90]  (2.2,1);

\draw[red,ultra thick] (1,1.33) -- (1.5,1.66);

\shade[ball color=blue] (0.5,0.33) circle (0.1);
\shade[ball color=blue] (1.5,0.33) circle (0.1);
\shade[ball color=blue] (1,0.66) circle (0.1);

\shade[ball color=blue] (1,1.33) circle (0.1);
\shade[ball color=blue] (1.5,1.66) circle (0.1);
\end{scope}

\begin{scope}[xshift=6cm]
\node[rounded corners=2pt, inner sep=4pt, fill=red!50] at (0,2) {c};
\draw (0.5,1) -- (0,0) -- (1, 0) -- (2,0) -- (1.5,1) -- (1,0) -- (0.5, 1) -- (1.5,1) -- (1,2) -- (0.5,1) ;

\draw[red,ultra thick] (0.5,0.33) -- (1,0.66);
\draw[red,ultra thick, dashed] (1.5,0.33) to[out=20,in=-90] (2.2,0.66);
\draw[red,ultra thick, dashed] (1,1.33) to[out=20,in=180] (1.7,1.66);

\shade[ball color=blue] (0.5,0.33) circle (0.1);
\shade[ball color=blue] (1.5,0.33) circle (0.1);
\shade[ball color=blue] (1,0.66) circle (0.1);
\shade[ball color=blue] (1,1.33) circle (0.1);
\end{scope}
\end{tikzpicture}
        \caption{Examples of dimer covering for an (a) even number of plaquettes compatible with plaquette matching, (b) odd number of plaquettes, (c) even number of plaquettes incompatible with plaquette matching. Solid lines are dimers on the dual lattice. Dashed lines indicate unmatched plaquettes.}
        \label{tikz:dimer_cover_Ex}
\end{figure}

\subsection{FKT algorithm}

Consider an undirected planar graph $G$ that describes the Ising degrees of freedom of a finite lattice with nearest-neighbor interactions. We can construct a dual graph $D$ of the interior of this lattice by taking the lattice plaquettes as vertices and adding edges if two plaquettes are adjacent. We call $A$ the adjacency matrix of this interior dual graph. We can employ the FKT algorithm for the graph $D$. For that we change the undirected graph $D$ into a directed graph $D'$ so that each existing edge becomes a directed edge. The FKT algorithm chooses a directed graph $D'$ that has a \textit{Pfaffian orientation}. If we do so the signed adjacency matrix $A'$ of $D'$ has a Pfaffian identical to the Hafnian of $A$. These conditions make the number of ground states on the original lattice computable as
\begin{equation}
    \# GS = 2  \operatorname {pf} (A')= 2 \sqrt{\det(A')}
\end{equation}
where $\operatorname {pf} (A')$ counts the number of dimer coverings on the dual and factor of $2$ because of $\mathbf{Z}_2$ symmetry was included. The ground-state entropy is hence given by
\begin{equation}
    S_{\rm res} = \ln(2)+\frac{1}{2} \ln ( \det A').
\end{equation}
This opens up the use of numerical approximations for the $\operatorname{logdet}$ to increase calculation speed.

\subsection{Dimer covering results for type-A systems}

In the case of an even number of plaquettes (arranged in a way that can host at least one dimer covering) the number of frustrated bonds in the ground state is exactly half of the number of plaquettes $N_P$. For an odd number we can connect one plaquette on the boundary to the ``outside'' and construct a dimer covering for the remaining lattice. The additional frustrated bond has $N_{bnd} \sim \gamma N$ possible positions on the boundary. For sufficiently large system sizes the residual entropy per spin should not change significantly by pinning the frustrated bond on the boundary. Therefore we can estimate that systems with odd numbered plaquettes have an additional entropy (per spin) contribution on the order of $s_{bnd} \sim \ln(\gamma N)/N \sim \mathcal{O}(\ln(N)/N)$. As this contribution vanishes in the large system limit we conclude that the family of type-A systems has a converging residual entropy per spin.
For dimer coverings (with at most one frustrated bond on the boundary) the ground-state energy per site is given as
\begin{equation}
 \frac{E_{GS}}{J \, N} = \frac{2 \lceil N_P / 2 \rceil - N_E}{N}.
\end{equation}
Using Eq.~\eqref{eq:euler-P} we can simplify this to
\begin{equation}
     \frac{E_{GS}}{J \, N} = \frac{1-N + (N_P\mod{2})}{N}.
     \label{eq:gs_energy}
\end{equation}
Using the closed expressions for $N$, $N_E$, and $N_P$ as function of layer number $n$ derived in Sec.~\ref{sec:hyper} for the $\{3,7\}$ system, we find the ground-state energy to be approximated by
\begin{equation}
    E_{GS}(n)/(J N) \approx  -1 + \frac{\sqrt{5}(1+\delta_n)}{7 \phi} \left( \frac{3+\sqrt{5}}{2} \right )^{-n},
\end{equation}
valid for large $n$. Here $\delta_n$ is one if $n$ is divisible by three and zero otherwise.


\subsection{Dimer covering results for type-B systems}

Remarkably, type-B lattices display a unique dimer covering for any lattice size. This can be shown inductively: Starting from the edge of an $n$-layer type-B lattice, each outermost plaquette has a single unique bond to the interior. Covering these bonds by dimers, the remaining lattice is a type-B lattice with $(n-1)$ layers. This reduction can be continued down to layer $0$.

As a result, the antiferromagnetic Ising model on a type-B lattice has a unique ground state (up to a global spin flip). We show below that this state is an ordered ferrimagnet.


\section{Monte-Carlo simulations}
\label{sec:mc}

Our Monte-Carlo simulations of the Ising model on hyperbolic tilings employ a standard metropolis algorithm with single-flip updates; this tends to freeze at low temperatures as expected from frustrated systems. To resolve this we utilize parallel tempering of multiple replicas, with replica temperatures chosen to have an acceptance ratio for replica switches of approximately 30\%.

\subsection{Thermodynamics}

\begin{figure}[!tb]
\includegraphics[width=\columnwidth]{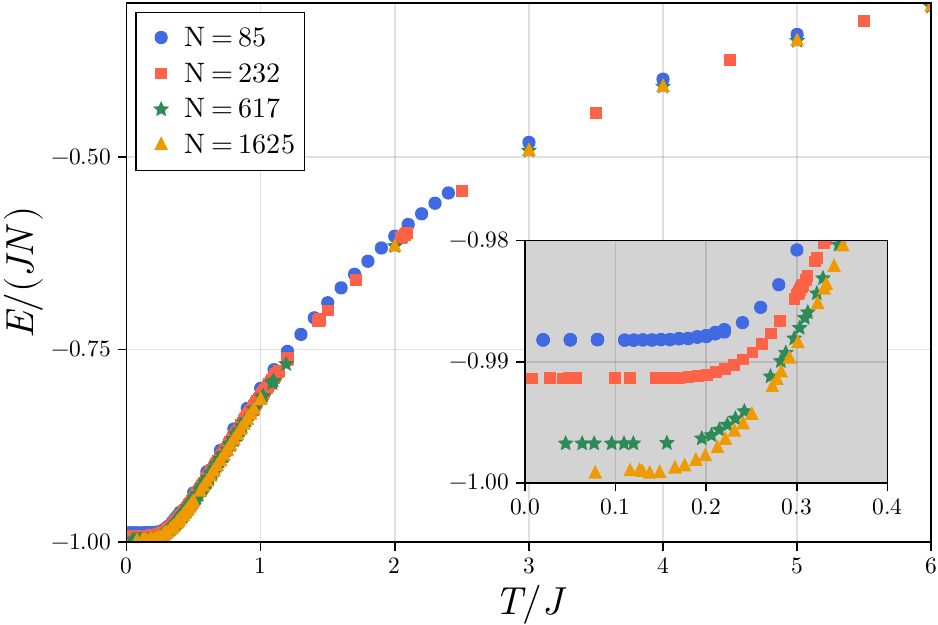}
\caption{
Energy per site, $E/N$, as function of temperature $T$ for $\{3,7\}$ type-A systems of different size $N$, obtained from MC simulations. The inset shows the low-temperature part of the data.
}
\label{fig:energy_vs_T}
\end{figure}

After equilibration, thermal averages are obtained from equally spaced samples along the MC run, with the spacing chosen according to the autocorrelation time \cite{Thomopoulos13}. The energy of the system (and other observables likewise) will be computed as
\begin{equation}
\langle E \rangle = \frac{1}{N_\tau}\sum_\tau E_\tau
\end{equation}
where $\tau$ runs over Monte Carlo time steps and $E_\tau$ is the system energy at that time step. We can infer from Fig.~\ref{fig:energy_vs_T} that the systems reaches the ground state energies predicted by Eq.~\eqref{eq:gs_energy} at low temperatures.
The heat capacity is obtained from
\begin{equation}
    C_V = \beta^2 \big(\langle E^2 \rangle - \langle E \rangle ^2\big)
\end{equation}
with $\beta=1/T$, and the uniform susceptibility according to
\begin{equation}
    \chi = \beta \big(\langle M^2 \rangle - \langle M \rangle ^2\big)
\end{equation}
which correspond to variances of the MC samples. To derive error estimates for the observables (esp. variances) we use the \textit{jackknife} rebinning method \cite{Thomopoulos13}.
The entropy is computed from integrating the heat capacity as described in the main text. At very low temperatures, $T\lesssim\Delta$ with $\Delta$ the gap between ground-state and lowest-excited-state energies, we employ
\begin{equation}
 S(T) \approx \ln{N_0} - \ln(\exp(\Delta \beta)+N_1) - \frac{\Delta \beta \exp(\Delta \beta)}{\exp(\Delta \beta)+N_1}
\end{equation}
where $N_{0,1}$ are the degeneracies of the ground state and the first excited state, respectively. The results of this fit are shown, e.g., in Fig.~2 of the main paper. With increasing sample size, statistical errors at low $T$ increase due insufficient sampling related to freezing.

\subsection{Resolution of spin correlations}
A spin-spin correlator $\sigma_i \sigma_j$ can only take values of $\pm 1$ in each individual measurement, while its thermodynamic average will be some finite value $\langle \sigma_i \sigma_j \rangle = \mu$. Repeated measurements of this observable are like repeated draws from a Bernoulli distribution
\begin{equation}
     P(\sigma_i \sigma_j = x)={\begin{cases}p&{\text{ if }}\quad x=+1\\1-p&{\text{ if }}\quad  x=-1\end{cases}}
\end{equation}
with $p$ depending on the model parameters. The measurements yield an expectation value $E[\sigma_i \sigma_j] = \mu = 2p-1$ and a variance $Var[\sigma_i \sigma_j] = 1-\mu^2$. The statistical error of the measured average is hence given by $\sigma_\mu = \sqrt{1-\mu^2}/\sqrt{D}$ where $D$ is the number of data points. An accurate measurement of a small $\mu$ therefore requires $D\gg1/\mu^2$. If spin correlations decay exponentially with distance, correlators at long distances, i.e., with very small $\mu$, cannot be measured accurately.


\section{Additional numerical results for thermodynamics}

We now exhibit numerical results from our MC simulations beyond those shown in the main paper.

\begin{figure}[tb]
\includegraphics[width=\columnwidth]{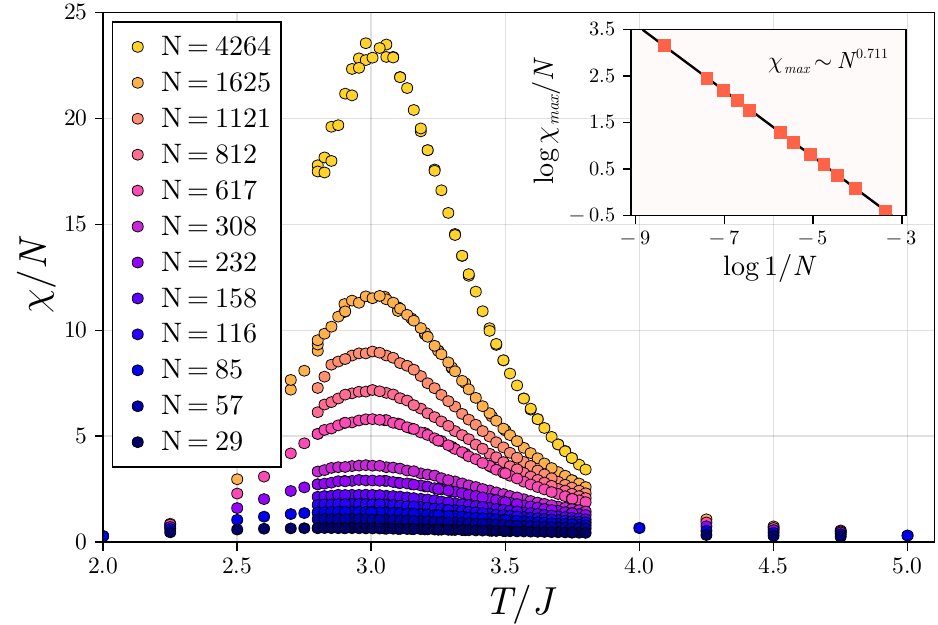}
\caption{
Uniform susceptibility $\chi = \beta \, var(|M|)$ versus temperature of the ferromagnetic Ising model on type-A $\{3,7\}$ systems of different sizes. Inset: Maximum value of $\chi(T)$ as function of system size $N$; the fit indicates a power-law divergence corresponding to a continuous phase transition.
}
\label{fig:suscep_absmag}
\end{figure}

\subsection{Ferromagnetic Ising model: Phase transition}

To benchmark the reliability of our MC simulations, we studied the \textit{ferromagnetic} Ising model on the same type-A $\{3,7\}$ tilings which yield spin-liquid behavior for the antiferromagnetic case. Here, we expect an ordered ferromagnetic ground state without any frustration effects. The natural order parameter is the uniform magnetization $M = \sum \sigma_i$. The corresponding susceptibility displays a large peak at a temperature of roughly $3J$ which sharpens (and slightly drifts) with system size, Fig.~\ref{fig:suscep_absmag}, indicating a finite-$T$ magnetic transition. In fact, the susceptibility maximum diverges with system size in a power-law fashion.

\begin{figure}[bt]
\includegraphics[width=\columnwidth]{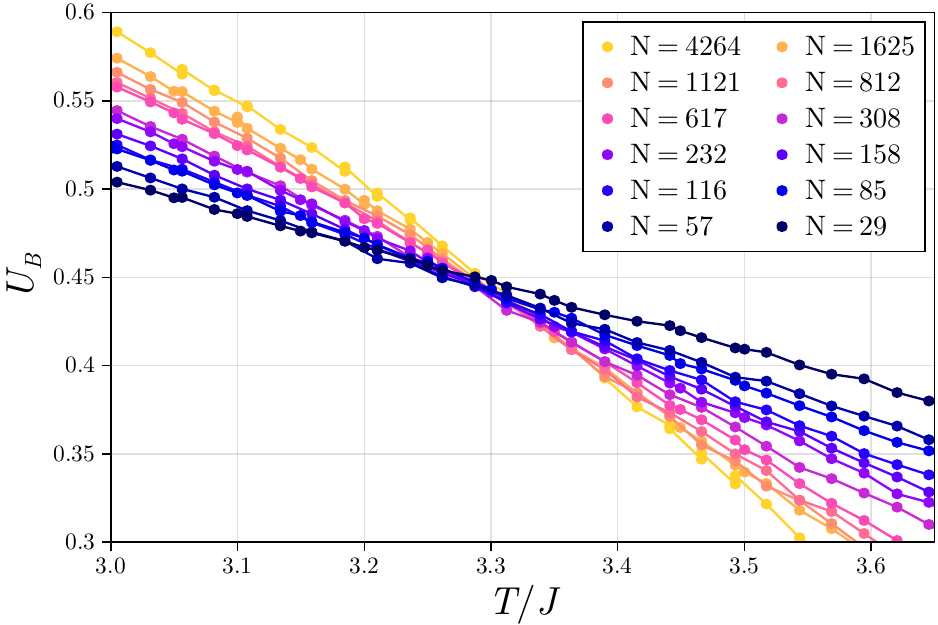}
\caption{
Binder cumulant of the uniform magnetization, $U_B(T)$, for the ferromagnetic Ising model.
The crossing points appear to converge to $T_c/J \approx 3.3$ with a non-trivial value of $U_B(T_c) = 0.45$.
}
\label{fig:UB_absmag}
\end{figure}

We confirm the phase transition by considering the binder cumulant
\begin{equation}
U_B=1-{\frac {{\langle M^{4}\rangle }}{3{\langle M^{2}\rangle }^{2}}}
\end{equation}
and observe a crossing of the data which extrapolates to the critical temperature $T_c \approx 3.3J$, see Fig.~\ref{fig:UB_absmag}.

\begin{figure}[bt]
\includegraphics[width=\columnwidth]{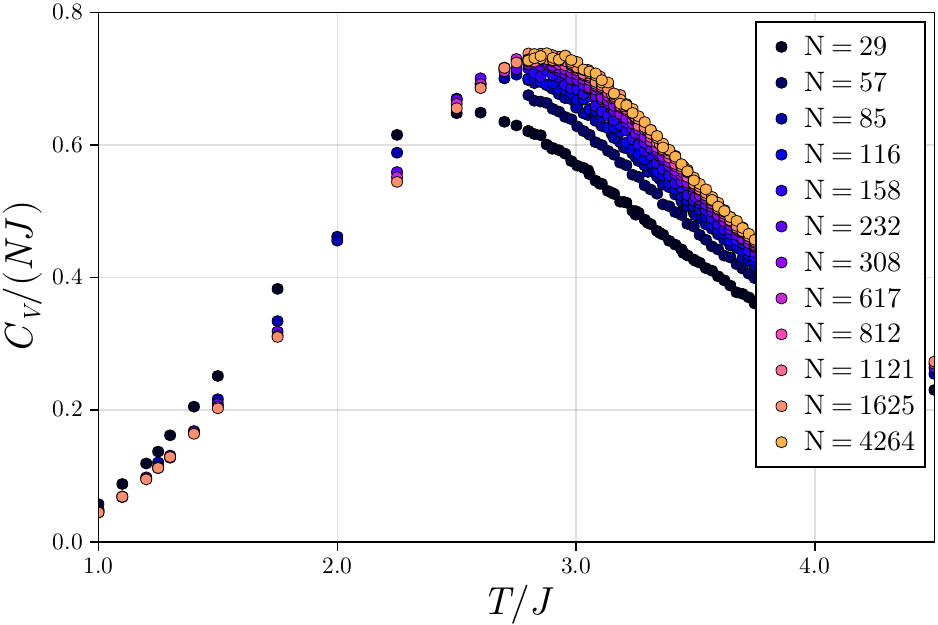}
\caption{Specific heat, $C_V(T)$, for the ferromagnetic Ising model.}
        \label{fig:FM_cv}
\end{figure}

We recall that standard finite-size scaling arguments for phase transitions, including the definition of critical exponents, need to be revisited for hyperbolic systems \cite{mnasri15} due to the exponential dependence between the total and the linear system size (i.e. radius). Contrary to flat space, the number of boundary sites of OBC systems approaches a fixed ratio of the total number of spins, see Sec.~\ref{sec:infla}. As a result, the system is effectively inhomogeneous and, in a certain sense, high-dimensional.
Consequent non-standard behavior is seen for instance in the specific heat, Fig.~\ref{fig:FM_cv}, which shows a pronounced bump near $T_c$, but no tendency to singular behavior. We note that MC simulations of the same model on hyperbolic lattices using periodic boundary conditions yielded a critical exponent $\alpha = 0$ \cite{breuckmann20}, reflecting mean-field-like behavior.

\subsection{Antiferromagnetic Ising model on type-B lattice: Ferrimagnet}

\begin{figure}[bt]
\includegraphics[width=\columnwidth]{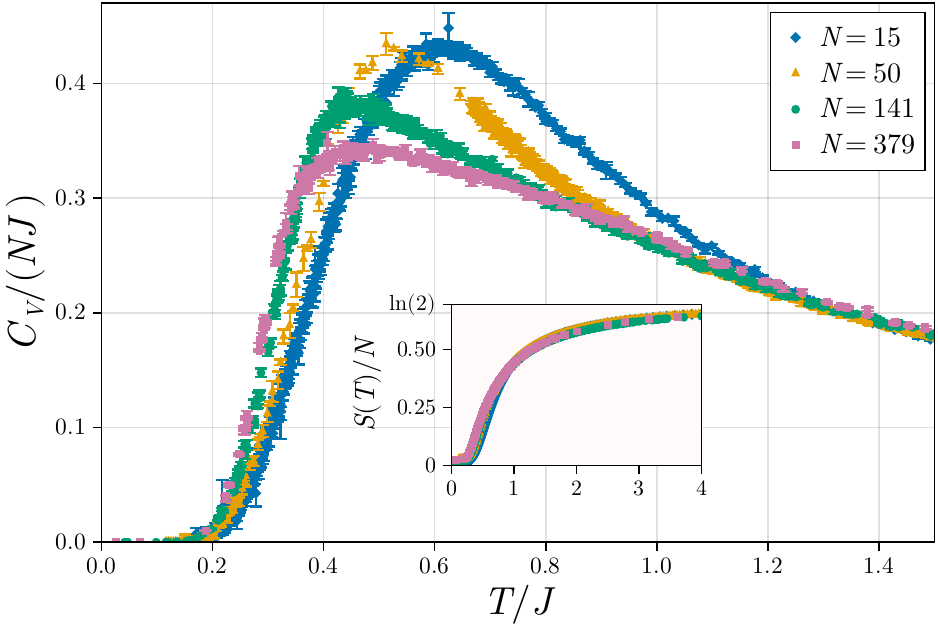}
\caption{
Specific heat (main) and entropy (inset) per site of the Ising antiferromagnet on type-B $\{3,7\}$ lattices of different size. The residual entropy vanishes within error bars. The data spread in $C_V(T)$ at intermediate temperatures reflects sampling errors, possibly related to a weak thermal hysteresis.
}
\label{fig:spiked_cv}
\end{figure}

In order to study the influence of the boundary shape on frustration effects, we study the antiferromagnetic Ising model on the type-B lattices introduced in Sec.~\ref{sec:typeblatt}. As noted above, the FKT algorithm signals a unique ground state (up to global spin flip). This is consistent with our MC results for the entropy, shown in the inset of Fig.~\ref{fig:spiked_cv}: The residual entropy per spin is zero.

The structure of the FKT dimer covering implies that the ground state consists of ferromagnetically ordered rings, with alternating orientation from ring to ring, i.e., a ferrimagnet. This is nicely seen in the NN spin correlations at low $T$ which we show in Fig.~\ref{fig:cooldown} below. In this state, all intralayer bonds are frustrated while all interlayer bonds are unfrustrated; given the construction of type-B lattices, the latter are the boundary bonds, and their large number tips the energetic balance towards this ordered state. We conclude that type-B systems display boundary-driven order.

\begin{figure}[tb]
\includegraphics[width=\columnwidth]{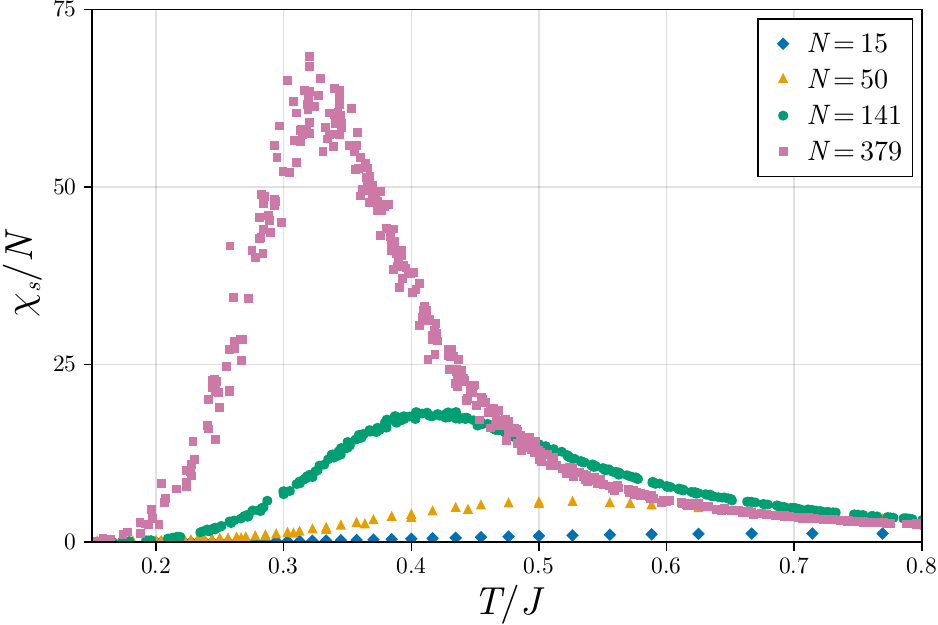}
\caption{Staggered susceptibility of the Ising antiferromagnet on type-B $\{3,7\}$ lattices of different size.}
\label{fig:spiked_stagchi}
\end{figure}

\begin{figure}[bt]
\includegraphics[width=\columnwidth]{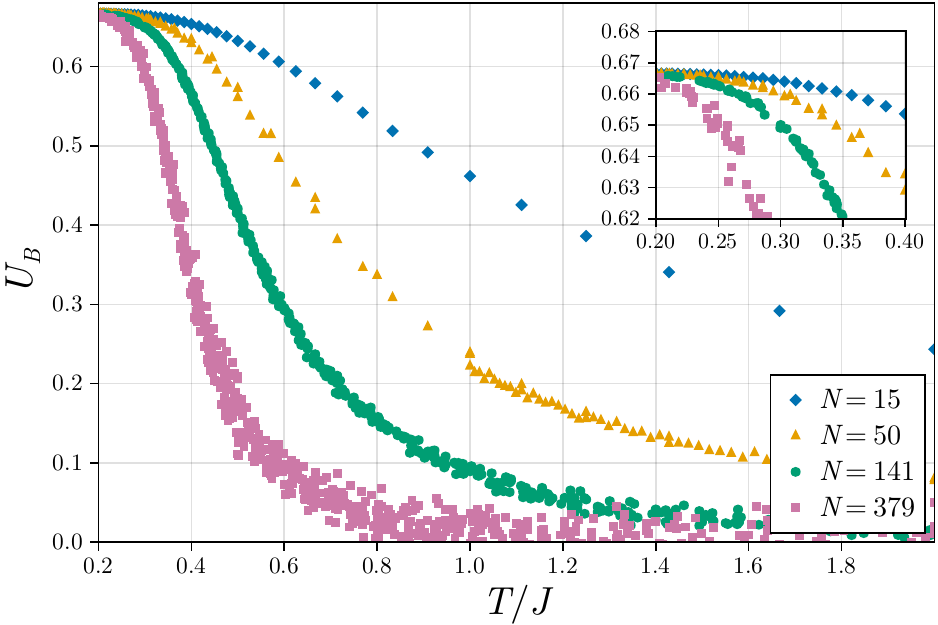}
\caption{
Binder cumulant of the staggered magnetization for type-B antiferromagnets. The inset shows a zoom into the low-$T$ data.
}
\label{fig:spiked_binder}
\end{figure}

The specific heat, Fig.~\ref{fig:spiked_cv}, is inconclusive with regards to the existence of a finite-$T$ transition. It features a relatively broad bump whose height appears to decrease with system size. We therefore consider the layer-staggered susceptibility $\chi_s$ which can be expected to diverge at the transition into the ferrimagnetic state. Indeed, the MC results in Fig.~\ref{fig:spiked_stagchi} display a significant maximum in $\chi_s(T)$ which increases with system size but also drifts significantly. The corresponding Binder cumulant is shown in Fig.~\ref{fig:spiked_binder}. It does not display a clear crossing point in the temperature regime where the peak in $\chi_s(T)$ is observed. We have also computed the Binder cumulant for the uniform magnetization which, however, does also not display a crossing point (not shown). Together, this suggests two possibilities for the thermodynamic behavior of type-B systems: Either the ordering temperature tends to zero (or an extremely small value) in the limit $N\to\infty$, or the transition into the ferrimagnet is of first order. We note that we have observed weak hysteretic behavior in our temperature sweeps, possibly consistent with a first-order transition, but we cannot exclude that the hysteresis is simply an effect of insufficient sampling. We therefore leave the precise study of this transition for future work.

We finally note that the disparate behavior of systems with type-A and type-B boundaries is a very special feature of hyperbolic OBC systems. It re-emphasizes the notion that the thermodynamic limit in hyperbolic (as opposed to flat) space cannot be discussed independent on boundary conditions \cite{ueda07,lux23,goetz24}.

\subsection{Antiferromagnetic Ising model on type-A lattice: Ground-state magnetization}

As discussed in the main text, the antiferromagnetic Ising model on type-A lattices features a highly degenerate ground state. The ground-state manifold encompasses states of different total magnetization; this leads to the appearance of a Curie contribution in the uniform susceptibility $\chi(T)$, as shown in Fig.~3 of the main paper.

\begin{figure}[tb]
\includegraphics[width=\columnwidth]{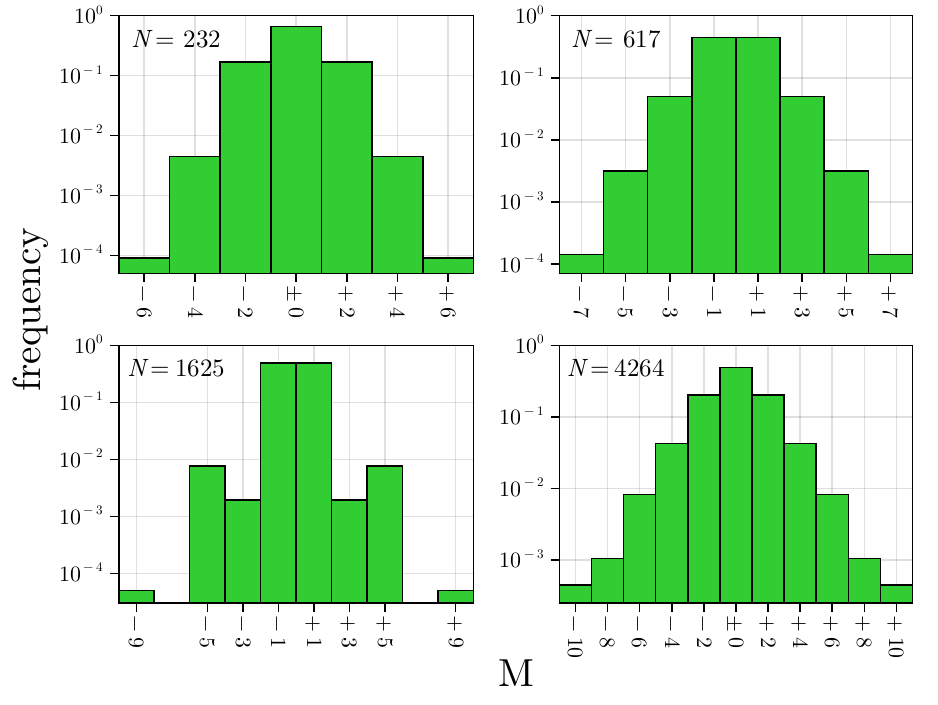}
\caption{
Distribution of the uniform magnetization in the ground-state manifold of the antiferromagnetic Ising model on type-A $\{3,7\}$ lattices of different size. Note that the vertical axis is logarithmic.
}
\label{fig:mdistr}
\end{figure}

We have determined the distribution of ground-state magnetizations for different system sizes by Monte-Carlo sampling, with results shown in Fig.~\ref{fig:mdistr}. While the distributions get wider with increasing system size, their width decreases slower than $\sqrt{N}$, such that the low-temperature Curie constant per spin, given by $\langle M^2 \rangle/N$, decreases with increasing $N$, consistent with Fig.~3 of the main paper.


\begin{figure}[!tb]
\includegraphics[width=0.9\columnwidth]{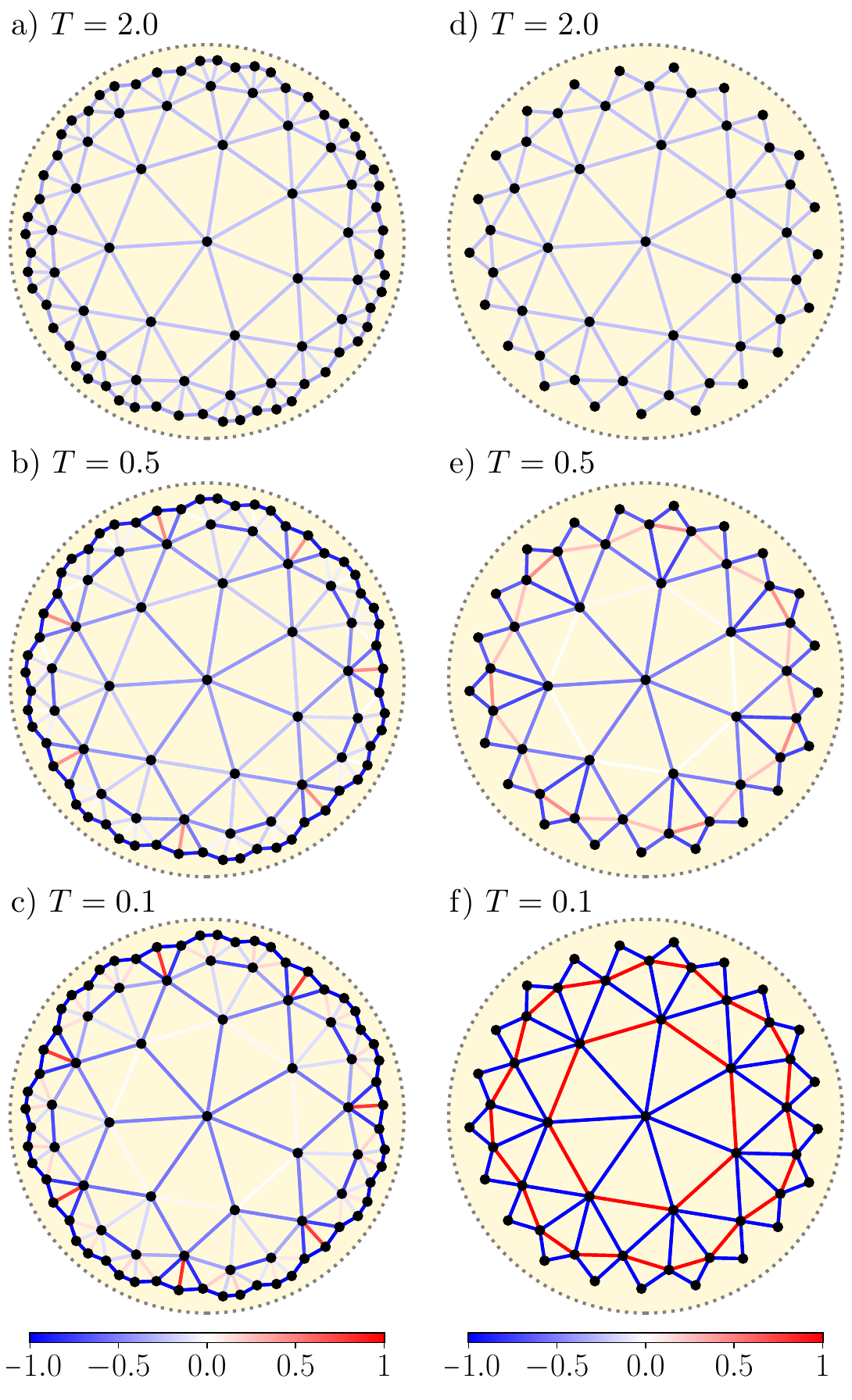}
\caption{
NN spin correlations, $\langle  \sigma_i \sigma_j \rangle$, for
(a-c) type-A $\{3,7\}$ lattices with $N=85$ sites and
(d-f) type-B $\{3,7\}$ lattices with $N=50$ sites.
(a) $T/J=2.0$, thermally disordered;
(b) $T/J=0.5$, with antiferromagnetic correlations along the boundary;
(c) $T/J=0.1$, spin-liquid regime;
(d) $T/J=2.0$, thermally disordered;
(e) $T/J=0.5$, with antiferromagnetic interlayer correlations;
(f) $T/J=0.1$, with ferromagnetic rings of the ferrimagnetic state.
}
\label{fig:cooldown}
\end{figure}

\section{Numerical results for correlation functions}

We now turn to spin-spin correlation functions obtained from our MC simulations.

\subsection{Nearest-neighbor correlations: type-A vs. type-B}

We start with an overview of thermally averaged nearest-neighbor correlators, complementing Fig.~7 of the main paper. Fig.~\ref{fig:cooldown} show a spatially resolved view of NN correlators for different temperatures for type-A and type-B systems: The differences in the low-$T$ correlations are obvious, reflecting spin-liquid and ferrimagnetic ground states, respectively.

\subsection{Distance-dependent intralayer correlations}

To further characterize the magnetic states, we compute spin correlations beyond nearest neighbors. Given the absence of simple distance definitions and corresponding Fourier transforms, we choose to analyze interlayer correlations. Each layer is a one-dimensional system which can be treated as a spin chain. Per layer we can compute the correlator
\begin{equation}
 C(\theta) = \frac{1}{N_\partial} \sum_\alpha \langle \sigma_\alpha \sigma_{\alpha+\theta} \rangle
\end{equation}

\begin{figure}[tb]
\includegraphics[width=\columnwidth]{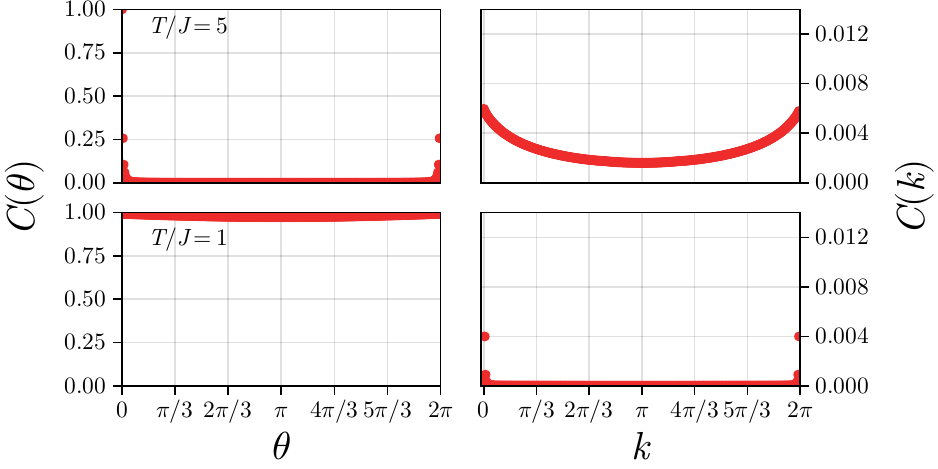}
\caption{
Boundary correlator $C(\theta)$ (left) and its Fourier transform (right) for a type-A $\{3,7\}$ system of size $N=385$ with ferromagnetic interactions.
Top: $T/J=5$, correlations decay exponentially;
Bottom: $T/J=1$, ferromagnetic order.
}
\label{fig:FM_edgecorr}
\end{figure}

\begin{figure}[tb]
\includegraphics[width=\columnwidth]{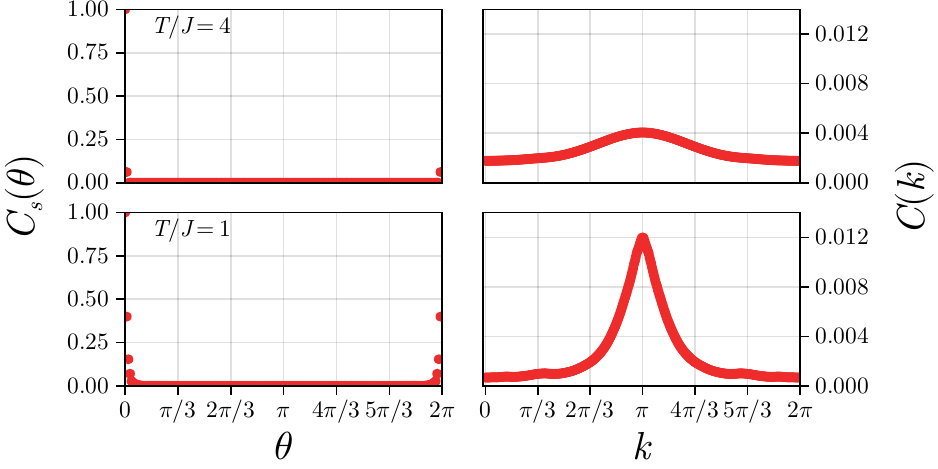}
\caption{
As in Fig.~\ref{fig:FM_edgecorr} but now for antiferromagnetic interactions, with the left panels showing the staggered correlator $C_s(\theta)=(-1)^{\theta N_\partial/(2\pi)} C(\theta)$.
Top: $T/J=4$, correlations decay exponentially, and $C(k)$ displays a weak maximum at $k=\pi$ reflecting antiferromagnetic correlations;
Bottom: $T/J=1$, the boundary approaches antiferromagnetic order.
}
\label{fig:AF_edgecorr}
\end{figure}

with $\theta = 2\pi (j-i)/N_\partial$, $i,j$ the (angular) spin positions in the layer, and $N_\partial$ the number of spins in the layer. $C(\theta)$ is by construction $2\pi$-periodic and can be Fourier-transformed into $C(k)$. At high $T$ and small distances, $C(\theta)$ decays exponentially $\sim \exp(-\theta N_\partial/ 2\pi \xi)$ where $\xi$ is a correlation length in units of the bond length.
We focus on correlations for the boundary layer.
\\
\newpage

\begin{figure}[H]
\includegraphics[width=\columnwidth]{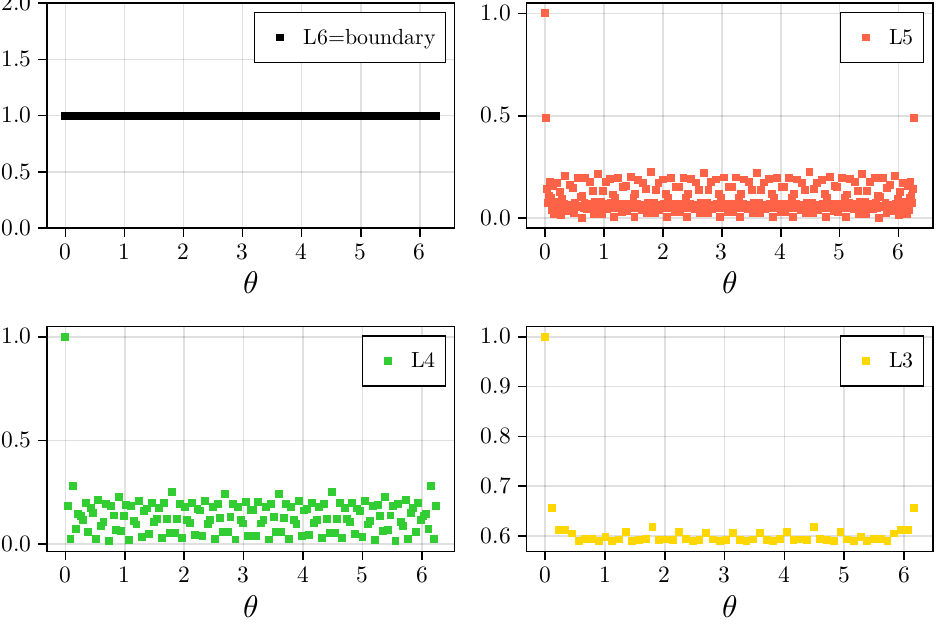}
\caption{
Boundary correlator $|C(\theta)|$ averaged over the ground-state manifold of the antiferromagnet with $N=1625$ sites. The four panels show the four outer layers $L_{3,4,5,6}$ of the six-layer system.
}
\label{fig:AF_layercorr}
\end{figure}
In case of ferromagnetic interactions the boundary orders ferromagnetically together with the bulk, and consequently the decay of $C(\theta)$ changes from exponentially decaying to long-range-ordered upon cooling, Fig.~\ref{fig:FM_edgecorr}.
For antiferromagnetic interactions the behavior is different. There is no bulk order at low $T$, as reflected by the extensive ground-state degeneracy, but the boundary tends to order antiferromagnetically at low $T$.
Due to frustration of the inner layers, correlations at the same temperature are generically weaker than for the ferromagnet, Fig.~\ref{fig:AF_edgecorr}. True order is only established at $T=0$, as there is no corresponding bulk order.
This has interesting consequences on the correlations in the inner layers. Those also decay exponentially at elevated $T$. Their low-$T$ behavior is illustrated in Fig.~\ref{fig:AF_layercorr} which shows an average over the ground-state manifold, i.e., for $T\to0$. While a short-distance decay is seen in all layers, correlations do not become small at long ``distances'' $\theta$. This is related to the hyperbolic geometry: Even spins which are apart by $\theta=\pi$ are connected by a rather short path through the center of the sample. Furthermore, the spread of the data reflects that the sites within each layer are not geometrically equivalent, but there is a structure with $q$-fold rotation symmetry ($q=7$ here).
